\documentclass[12pt,preprint]{aastex}
\usepackage{natbib,graphicx}
\bibpunct{(}{)}{;}{a}{}{,}

\def\ie {i.e.}
\def\eg {e.g.}

\shorttitle{A Radio Study of Seyfert galaxy Mrk~6}
\shortauthors{Kharb, et~al.}

\begin{document}

\title{A Radio Study of the Seyfert galaxy Markarian~6: Implications for
Seyfert life-cycles}
\author{P. Kharb}
\affil{Centre for Imaging Science, Rochester Institute of 
Technology, 54 Lomb Memorial Drive, Rochester, NY 14623}
\email{kharb@cis.rit.edu}
\author{C. P. O'~Dea}
\affil{Dept. of Physics, Rochester Institute of Technology, 
84 Lomb Memorial Drive, Rochester, NY 14623}
\author{S. A. Baum}
\affil{Centre for Imaging Science, Rochester Institute of Technology, 
54 Lomb Memorial Drive, Rochester, NY 14623}
\author{E. J. M. Colbert}
\affil{Johns Hopkins University, Dept. of Physics and Astronomy,
3400 North Charles Street, Baltimore, MD 21218}
\author{C. Xu}
\affil{Space Telescope Science Institute, 3700 San Martin Drive, 
Baltimore, MD 21218}

\begin{abstract}
We have carried out an extensive radio study with the Very Large Array on 
the Seyfert 1.5
galaxy Mrk~6 and imaged a spectacular radio structure in the source. The radio 
emission occurs on three different spatial scales, from $\sim$7.5~kpc bubbles 
to $\sim$1.5~kpc bubbles lying nearly orthogonal to them and a $\sim$1~kpc 
radio jet lying orthogonal to the kpc-scale bubble. To explain the 
complex morphology, we first consider a scenario in which the radio 
structures are the result of superwinds ejected by a nuclear starburst.
However, recent Spitzer observations of Mrk~6 provide an upper limit to the 
star formation rate (SFR) of $\sim5.5~M_\sun$~yr$^{-1}$, an estimate much 
lower than the SFR of $\sim33~M_\sun$~yr$^{-1}$ derived assuming that the 
bubbles are a result of starburst winds energized by supernovae explosions.
Thus, a starburst alone cannot meet the energy requirements for the creation 
of the bubbles in Mrk~6.
We show that a single plasmon model is energetically infeasible, and we
argue that a jet-driven bubble model while energetically feasible does
not produce the complex radio morphologies.
Finally, we consider a model in which the complex radio structure is a
result of an episodically-powered precessing jet that changes its
orientation. This model is the most attractive as it can naturally explain the 
complex radio morphology, and is consistent with the energetics, the
spectral index and the polarization structure. Radio emission in this scenario 
is a short-lived phenomenon in the lifetime of a Seyfert galaxy which 
results due to an accretion event. 
\end{abstract}


\section{INTRODUCTION}

Radio observations show that most Seyfert galaxies have sub-parsec-scale
radio emission 
\citep[e.g.,][]{deBruyn76,UlvestadWilson84A,UlvestadWilson84B,Roy94,
Thean00,Lal04}.
Some of these galaxies with small-scale radio emission show elongated 
structures, similar to the radio jets seen in powerful radio galaxies. 
Radio emission on kiloparsec scales has also been detected in some Seyfert 
galaxies. The large-scale radio structures however are not typically found 
to be aligned with the small scale jets \citep{Baum93,Colbert96,Gallimore06}.
Further, Seyfert radio jets were found to be randomly aligned with respect
to the host galaxy major axes \citep{Kinney00,Schmitt01,SchmittKinney02}.
Abrupt changes of jet axes were also found in some Seyfert 
galaxies, e.g., NGC~4151 \citep{Ulvestad98} and NGC~1068 \citep{Gallimore96}.
It appears that the radio structures in Seyfert galaxis are more complex
than in the radio galaxies. Extensive discussions of large scale radio 
structures in Seyfert galaxies can be found in \citet{Baum93,Colbert96} and 
\citet{Gallimore06}.

Mrk~6 is an early-type (S0a) Seyfert 1.5 galaxy. It is also
one of the Seyfert galaxies that have radio emission on both small and
large scales. Using the Westerbork Synthesis Radio telescope (WSRT), 
\citet{Baum93} detected a pair of large weak ``radio lobes" extending 
northeast-southwest and a smaller scale radio structure extending east-west. 
High resolution MERLIN observations of the nuclear region by \citet{Kukula96} 
revealed a well-defined jet extending in the north-south direction. 
This jet is nearly aligned with the ionization cone reported earlier 
by \citet{Meaburn89}, and the optical polarization position angle (P.A.) 
derived through spectropolarimetric observations \citep{Smith04}. 
\citet{Kukula96} also found some evidence of a pair of 
lobes on the sides of the jet which corresponded to the east-west radio 
structure observed by \citet{Baum93} and \citet{Nagar99}. 
However, due to the relatively poor resolution of
WSRT ($\sim3\arcsec.5$) and the limited $u$--$v$ coverage of MERLIN,
few conclusions could be made on the relation between the
three scales of radio emission in this source. In order to have a better
understanding of the radio structures in Mrk 6, we undertook a comprehensive
radio study of Mrk~6 with the Very Large Array (VLA).

Throughout this paper we assume 
$H_0$=71 km~s$^{-1}$Mpc$^{-1}$, $\Omega_{m}=0.27$ and $\Omega_{vac}=0.73$.
Therefore, at the redshift of $z$=0.01881 (recession velocity 
$\sim$5640~km~sec$^{-1}$), or at the distance of 80.6 Mpc for Mrk~6,
1~arcsec is equivalent to 377~parsecs.

\section{OBSERVATIONS AND DATA ANALYSIS}

Mrk~6 was observed with the VLA at $\lambda$~6~cm (4860 MHz)
and $\lambda$~20~cm (1430 MHz) during 1995\---1996. Almost all
VLA configurations were used in the observations in order to 
image Mrk~6 on different spatial scales. 3C~286 was used as the primary 
flux density and polarization calibrator, while 0614+607 was used as the 
phase calibrator for 
the entire experiment. The data were processed with NRAO's software package 
AIPS using the standard imaging and self-calibration procedures. 
Table~\ref{tabobs} lists the observing frequency, the corresponding bandwidth, 
the VLA configuration, the observation date, the FWHMs of the synthesized beams,
the total flux densities of the source at various resolutions,
and the {\it r.m.s.} noise in the final maps.
\clearpage
\begin{table}[h]
\begin{center}
\caption{\label{tabobs} Observation Log}
\begin{tabular}{cccccccc} \hline \hline
Frequency&BW&Config.&Observation & Resolution&Total Flux Density&r.m.s. \\
(GHz)&(MHz)&& Date&(arcsec$^2$)&(mJy)&($\mu$Jy/bm) \\
\hline 
4.86&50&A&07/21/1995&0.41$\times$0.31&85.8&16 \\
4.86&50&B&11/02/1995&1.43$\times$1.13&96.4&16 \\
4.86&50&C&02/17/1996&4.27$\times$4.04&96.5&25 \\
4.86&50&D&05/03/1995&16.2$\times$11.4&100.5&29 \\
1.43&50&A&07/03/1995&1.44$\times$1.12&274.4&27 \\
1.43&50&C&02/17/1996&13.9$\times$13.0&270.8&147 \\
1.43&50&D&06/01/1995&59.1$\times$38.2&275.0&210 \\
&&&& \\ \hline \\
\end{tabular}
\end{center}
\end{table}
\clearpage
\section{RESULTS}
The total intensity radio maps are presented in Fig.~\ref{figmorph} while
Fig.~\ref{figwfpc} shows the relative orientation of the radio structures
with respect to the extended emission-line region (ENLR) 
and host galaxy of Mrk~6.
Figures~\ref{figpoln} and \ref{figspectr} show the radio polarization and the
spectral index images, respectively.
We describe below the primary findings of our study.

\subsection{Radio Morphology: Emission on Three Spatial Scales}
\begin{figure}
\includegraphics[width=16cm]{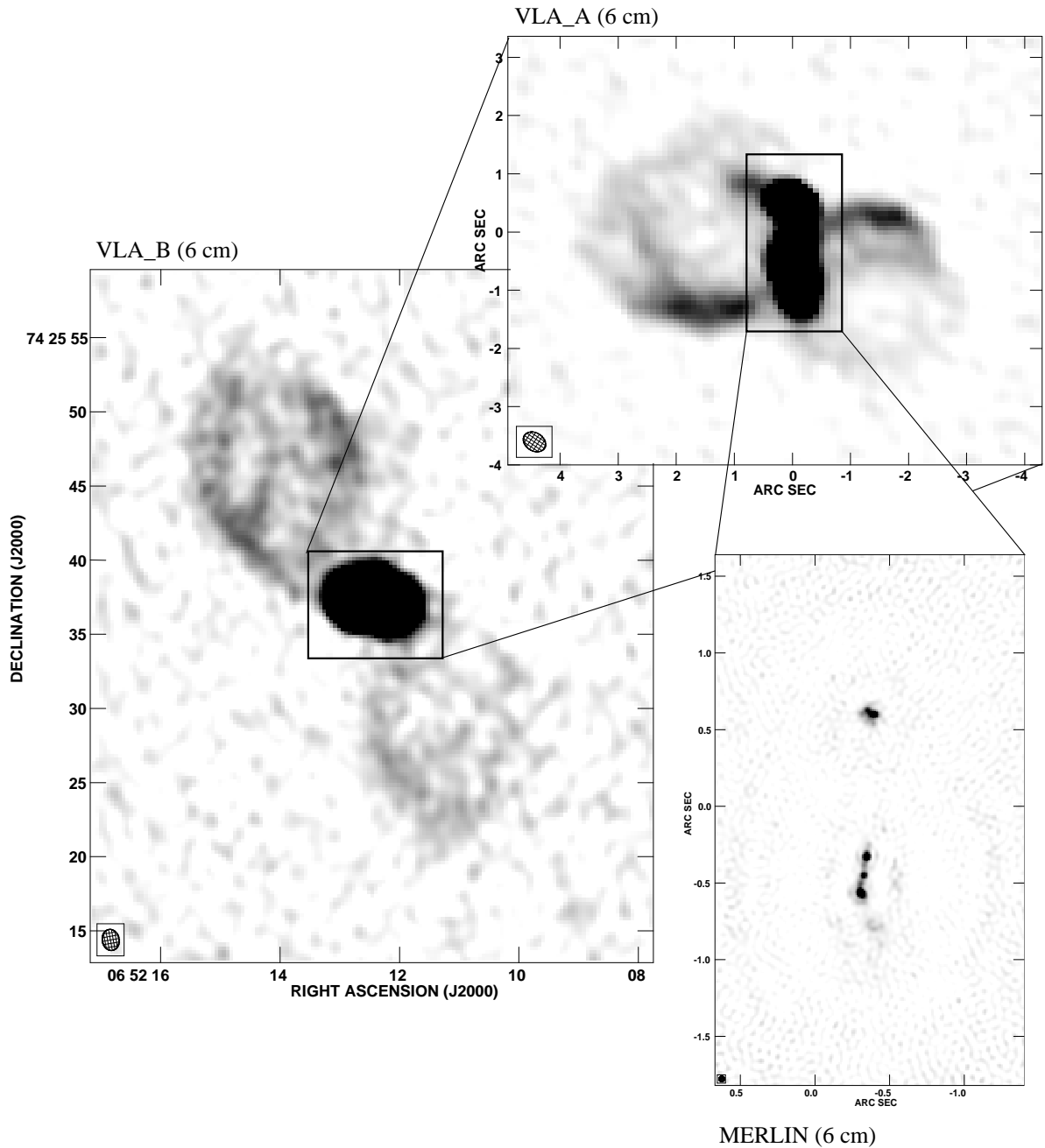}
\caption{Grey scale images of the 6~cm radio structures observed in Mrk 6.
The bubble-like structures are clearly seen in the VLA images.
The MERLIN map of the inner jet is taken from \citet{Kukula96}.
The centre of the optical host galaxy lies close to the (0,0) position in 
the MERLIN image, with a $3\sigma$ uncertainity of $0.27\arcsec$ 
\citep{Clements81}. There is no radio `core' coincident with this position.}
\label{figmorph}
\end{figure}

\begin{figure}
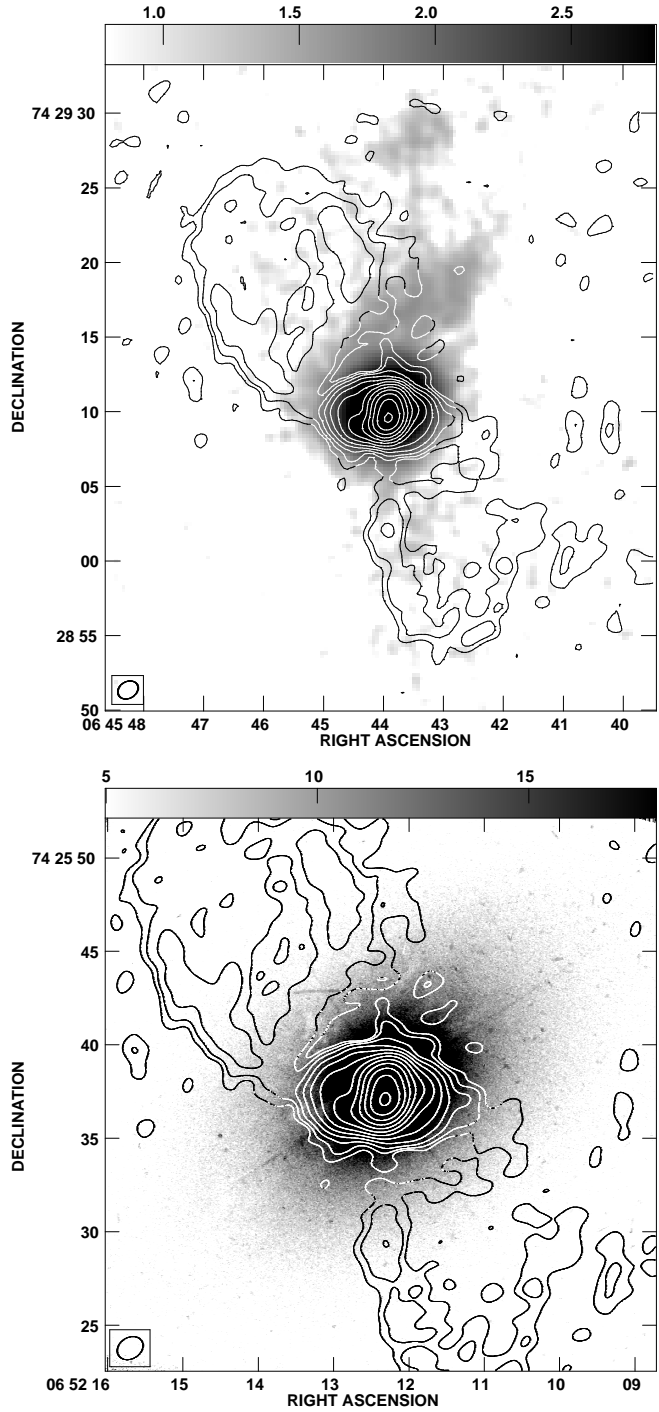

\centering{
\includegraphics[width=9cm]{f2a.ps}
\includegraphics[width=9cm]{f2b.ps}}
\caption{The 20~cm radio contours superimposed
on the grey-scale images of (top) extended [OIII] emission-line region 
\citep{Kukula96}, and (bottom) HST (WFPC2/F606W) optical continuum from the 
host galaxy (obtained from the HST archive). The outer bubble-like 
structures are projected roughly orthogonal to the major axis of the 
bulge-dominated S0a host galaxy. However, the structures lie within the
extent of the host galaxy.}
\label{figwfpc}
\end{figure}

Figure~\ref{figmorph} displays the mosaic of radio emission at 6 cm. The radio 
map of the inner jets is from the MERLIN observations of \citet{Kukula96}.
It is immediately apparent that Mrk 6 has radio emission on three scales.
First, there is a  radio jet of extent $\sim$1 kpc (3${\arcsec}$) 
extending in the north-south direction, at the active galactic nucleus (AGN). 
Second, there are a pair of inner bubbles of size $\sim$1.5$~\times~$1.5~kpc 
($\sim4\arcsec\times4\arcsec$) each, extending east-west and nearly 
perpendicular to the central jets. And third, there are a pair of outer 
bubbles of dimensions $\sim$4.5$~\times~$7.5~kpc 
($\sim12\arcsec\times20\arcsec$) each, extending northeast-southwest. 
In Fig.~\ref{figwfpc} we observe the relative orientation of these radio
structures with respect to the extended [OIII] emission-line region as observed
by \citet{Kukula96} and the host galaxy optical continuum from the 
high-resolution HST image (obtained from the HST archive). 
We note that the inner kpc jet is roughly
aligned with the emission-line gas. Further, the outer bubbles are
aligned with the host galaxy minor axis. The SDSS image from the
NASA/IPAC Extragalactic Database (NED)
indicates that the extent of the galaxy is $>15\times20$~kpc; the 
radio structures therefore lie within the confines of the host galaxy. 
\clearpage
\clearpage
We have listed some geometric and physical parameters pertaining to these 
individual radio structures, including their 
sizes, position angles, flux density and spectral indices in 
Table~\ref{tabprop}.
For the outer pair of bubbles, the radio flux density of the northeast 
bubble is about two and a half times that of the southwest bubble.
For the inner pair, the radio flux density of the east bubble is about two 
times that of the west bubble. 
We note that (1) the two pairs of bubbles and the jets show 
self-similar characteristics in their morphologies and sizes, \ie, 
the width of the larger structure corresponds to the overall size of
the next smaller structure, (2) the bubbles show clear edge-brightening and
(3) the bubbles contain filamentary structures. 
\clearpage
\clearpage 
\begin{table}[h]
\begin{center}
\caption{\label{tabprop} Radio Properties of Mrk~6}
\begin{tabular}{rcccccccccccccc} \hline \hline
Structure&Size&P.A.& S$_{6}$&L$_{6}$&$\alpha_6^{20}$\\
&(arcsec$^2$) &&(mJy)&(W~Hz$^{-1}$)&\\
\hline
Jet&1$\times$3&179$\arcdeg$&75.0& 5.8E22 &$-0.91\pm$0.21\\
North &...&...&...&...&$-0.90\pm$0.14\\
South &...&...&...&...&$-0.84\pm$0.10\\
Inner Bubble&...&76$\arcdeg$&11.4& 8.9E21 &...\\
East &4$\times$4&...&7.9& 6.1E21 &$-0.61\pm$0.37\\
West &3$\times$3&...&3.5& 2.7E21 &$-0.55\pm$0.30\\
Outer Bubble&-&33$\arcdeg$&9.4& 7.3E21&...\\
North &12$\times$20&-&6.2& 4.8E21&$-0.71\pm$0.21\\
South &12$\times$18&-&3.2& 2.5E21&$-0.55\pm$0.26\\
\hline 
\end{tabular}
\end{center}
\tablecomments{Col.2: Extents of structures using the 6~cm maps;
Col.4: 6~cm flux density in mJy; Col.5: 6~cm luminosity in W~Hz$^{-1}$;
Col.6: Spectral index estimated using the 6~cm and 20~cm images.}
\end{table}
\clearpage
\subsection{Radio Polarization}

\begin{figure}
\centering{
\includegraphics[width=12cm]{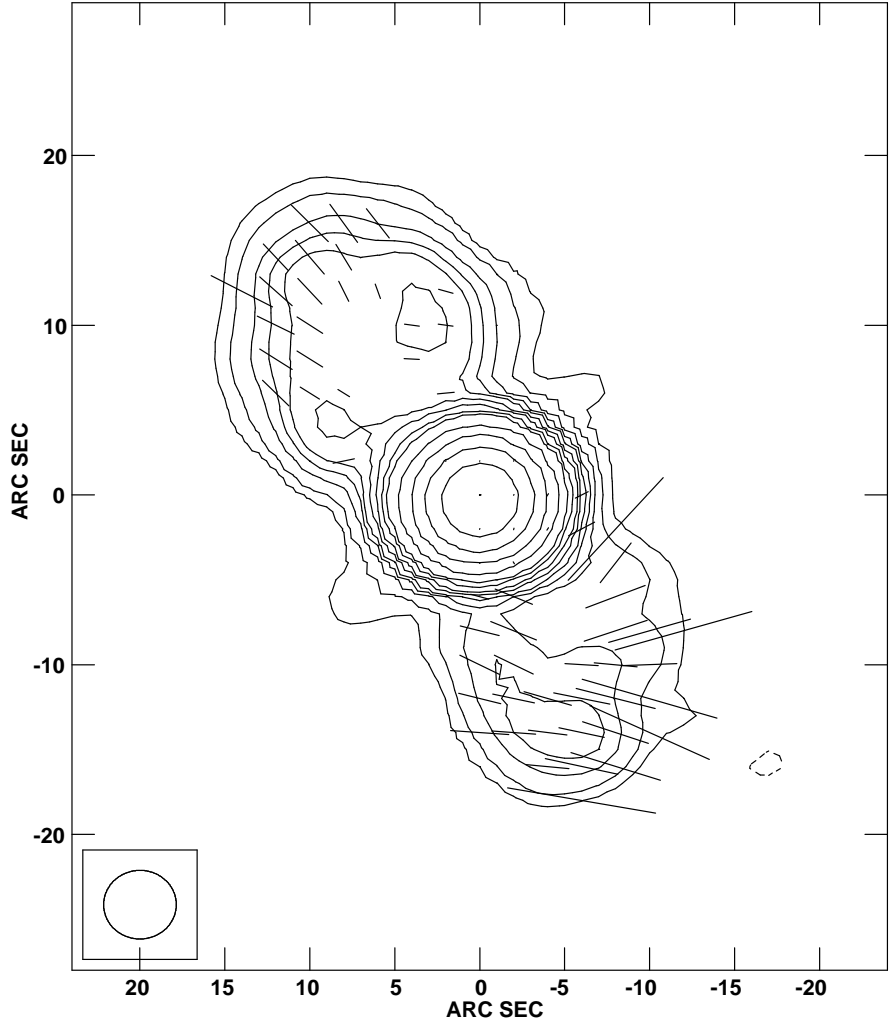}}
\caption{Contour image of Mrk~6 at 6~cm with polarization electric vectors 
and percentage of polarization indicated. The image was made using the
VLA C-configuration data. The contour levels are $7.45~10^{-5} \times$ 
(-1,1,2,4,6,8,12,16,24,32,64,128,
256,512) Jy/beam, with the beam size $4\arcsec.2 \times 4\arcsec.0$.
The lowest contour level corresponds to 3 times the off source r.m.s noise.
1$\arcsec$ in length of polarization vector corresponds to 10\% of polarization.} 
\label{figpoln}
\end{figure}

Polarization was detected at 6 cm in the region corresponding to the
western edge of the inner bubble and on the edges of the outer pair of 
bubbles. Figure~\ref{figpoln} displays the 6 cm VLA C-configuration 
image with polarization electric vectors superimposed. 
The length of the vectors is proportional to the fractional polarization.
At this configuration, polarization was detected at the 10--20 $\sigma$ 
levels (or 0.2--0.4 mJy~beam$^{-1}$). This corresponds to a high fractional 
polarization of $\sim 50\%$ at the edges and less than
1\% polarization near the center. 

We were unable to detect polarization
on the edge of the outer bubbles with the B-configuration, and anywhere with
A-configuration, due to the smaller beam size and lower resultant surface
brightness sensitivity. A suitable taper on the B-configuration $uv$-coverage 
recovers the polarization features observed in the C-configuration image 
(Fig.~\ref{figpoln}), thus confirming the detection 
of polarization. Assuming that the radio emitting medium is optically thin, 
and the Faraday rotation is small at 6~cm \citep[Galactic rotation measure 
is $\le30$~rad~m$^{-2}$ in the direction of Mrk~6,][]{SimardNormandin80}, 
we infer the magnetic field orientation to be mostly aligned with the edge of 
the bubbles. 
\clearpage
\clearpage
\subsection{Spectral Indices}

\begin{figure}
\centering{
\includegraphics[width=8.2cm]{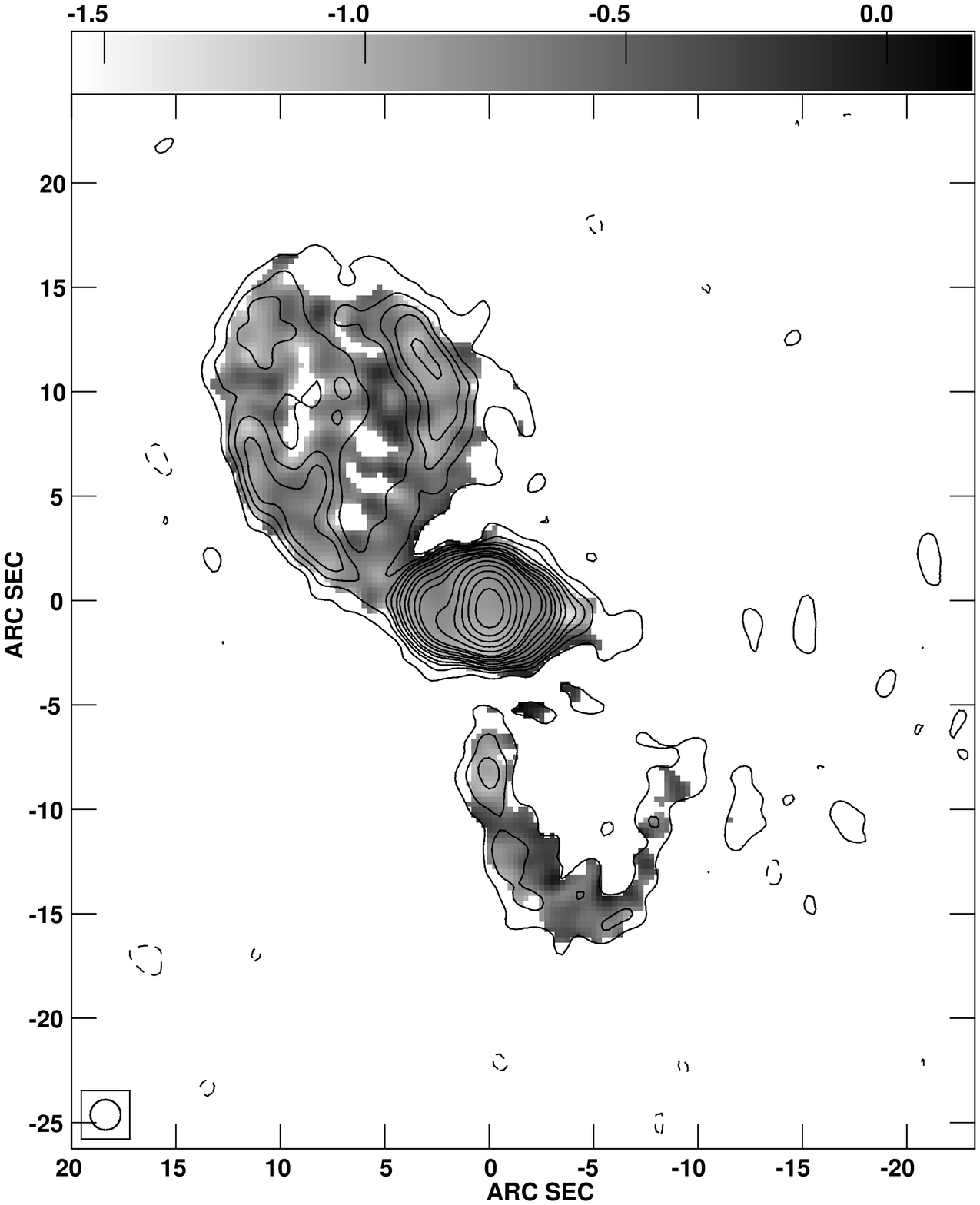}
\includegraphics[width=8.2cm]{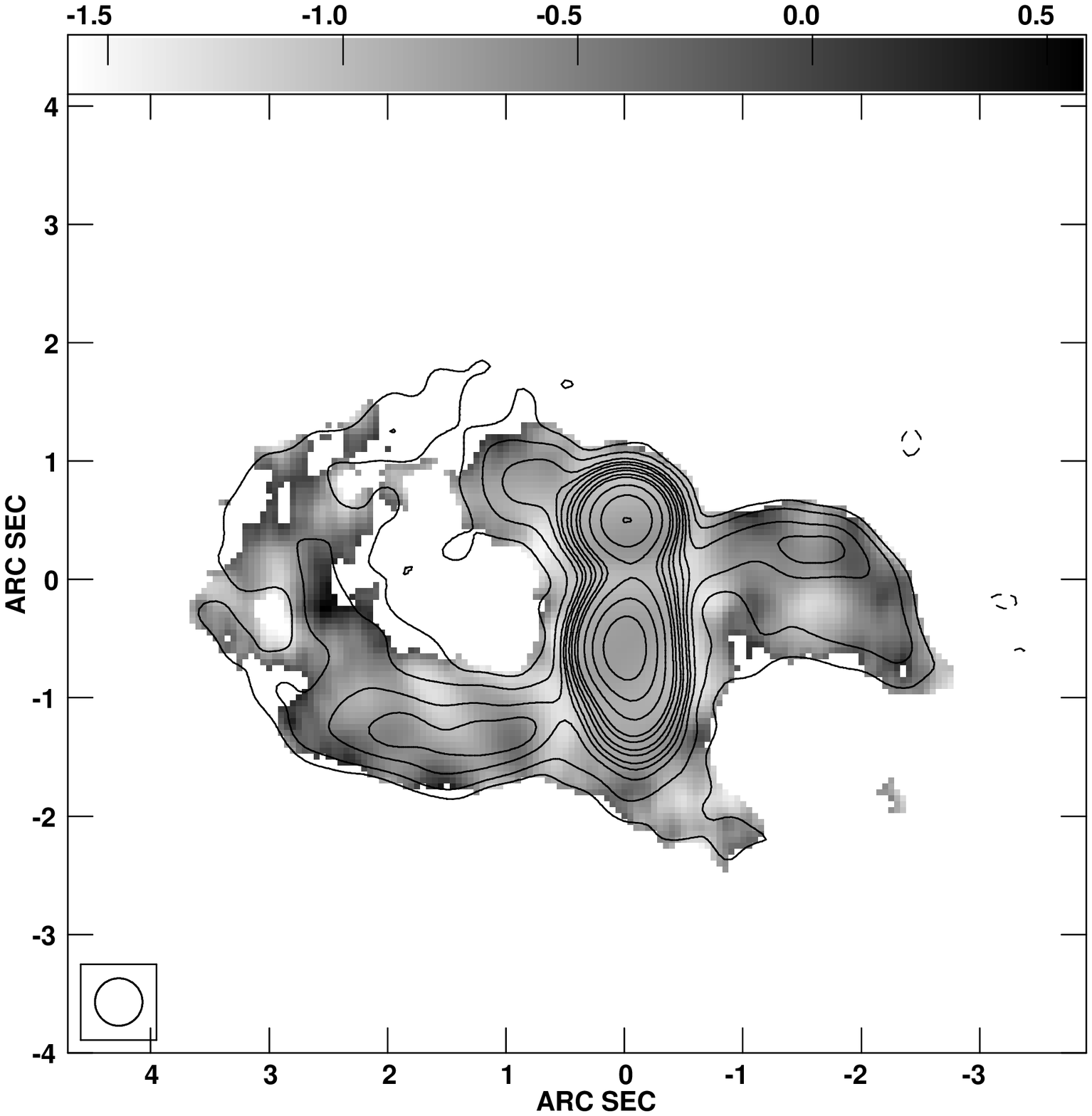}}
\caption{Spectral index map (see text) in greyscale overlaid on 
the (top) 20~cm A-configuration and (bottom) 6~cm A-configuration contour maps. 
The contour levels and beamsizes are (top) $8.82~10^{-5} 
\times$ (-1,1,2,4,6,8,12,16,24,32,64,128,256,512) Jy/beam and 
$4\arcsec.5 \times 4\arcsec.5$ and (bottom)
$7.0~10^{-5}\times$ (-1,1,2,4,6,8,12,16,24,32,64,128,256,512) Jy/beam
and $0\arcsec.5 \times 0\arcsec.5$.}
\label{figspectr}
\end{figure}

Figure~\ref{figspectr}~(top) shows the 6 cm B-configuration contour image 
overlaid on the grey scale spectral index image, which was made from the 6~cm 
B-configuration and the 20~cm A-configuration images, after restoring both
images with the same synthesized beam. The spectral indices vary only 
slightly across the image. The average spectral index over the central 
region corresponding to the inner bubbles is $-0.78\pm$0.26, that over the 
northern bubble is $-0.71\pm$0.21 and that over the southern bubble is 
$-0.55\pm$0.26. These spectral indices are consistent given the large errors.

Figure~\ref{figspectr}~(bottom) displays the spectral index map between 6~cm 
and 18~cm, in the neighbourhood of the jets. This image was made by combining 
our 6~cm A-configuration map with the 18~cm MERLIN map from \citet{Kukula96},
at the matched resolution of $0\arcsec.4\times0\arcsec.4$. It shows that the 
spectral index of the northern hotspot in the jet is $-0.90\pm$0.14, that of 
the southern hotspot is $-0.84\pm$0.10 and that of the saddle between the two 
hotspots is $-0.96\pm$0.08. Our results are in general agreement with those 
of \citet{Kukula96}. The average spectral index over the eastern bubble 
is $-0.61\pm$0.37 and over the western bubble is $-0.55\pm$0.30, slightly 
flatter than those in the jets, but similar to those in the outer pair of 
bubbles. 

\subsection{Energetics}
\label{energetics}

We have estimated the magnetic field strengths in regions corresponding to the
bubbles assuming equipartition
of energy between relativistic particles and the magnetic field
\citep{Burbidge59}. Using Eqns. 1--5 of \citet{OdeaOwen87}, we
have also obtained the minimum pressure and particle energy (electrons 
and protons) at minimum pressure. These estimates are listed in 
Table~\ref{tabenergetics}. The averaged total radio luminosity (L$_{rad}$) 
was estimated assuming that the radio spectrum extends from 10~MHz to 10~GHz 
with a spectral index of $\alpha$=--0.6. Further, it was assumed that the 
relativistic protons and electrons have equal energies. 

In Table~\ref{tabenergetics} we list
the minimum pressure values for the two scenarios where the bubbles are
completely filled (volume filling factor, $\phi$=1) or are mostly empty 
($\phi$=10$^{-3}$). 
The estimated minimum pressure in the filled outer bubbles turns
out to be $\sim10^{-12}$~dynes~cm$^{-2}$
and in the inner bubbles P$_{min}$ is $\sim10^{-11}$~dynes~cm$^{-2}$. 
The pressure is about fifty times larger for the bubbles when the
volume filling factor is $\phi$=10$^{-3}$. 
\clearpage
\clearpage
\section{DISCUSSION}

Bubbles and rings have been observed in a few radio galaxies and quasars, for 
example, in Hercules A \citep{Dreher84,Baum96}, 3C~310 \citep{vanBreugel84}, and
Centaurus A \citep{Quillen06}. Bubble-like structures have also been observed 
in some Seyfert and starburst galaxies, \eg, NGC~2992, NGC~3079, NGC~5548, 
NGC~4051, and NGC~6764 \citep{WehrleMorris88,Duric88,Baum93,Nagar99}, 
and planetary nebulae such as $\eta$ Carina and Hubble 5 
\citep{Nota95,Dwarkadas98}. Several mechanisms have been suggested for 
the formation of these bubble-like structures. We describe them in 
Sect.~\ref{secmech} and subsequently examine three potentially viable
models for the formation of the radio structures in Mrk~6.

\subsection{Constraints On A Model For Mrk~6}

We note that the bubble-like structures observed in the above mentioned 
radio galaxies are essentially spherical and they are either part of or 
directly connected with the radio jet. This is different from the 
bubbles observed in Mrk~6 which are elongated, larger than the jet and 
not aligned with it. 
The elongated morphology of the bubbles make them more similar to 
the bubbles observed in planetary nebulae.   

\begin{figure}
\centerline{
\includegraphics[width=11cm]{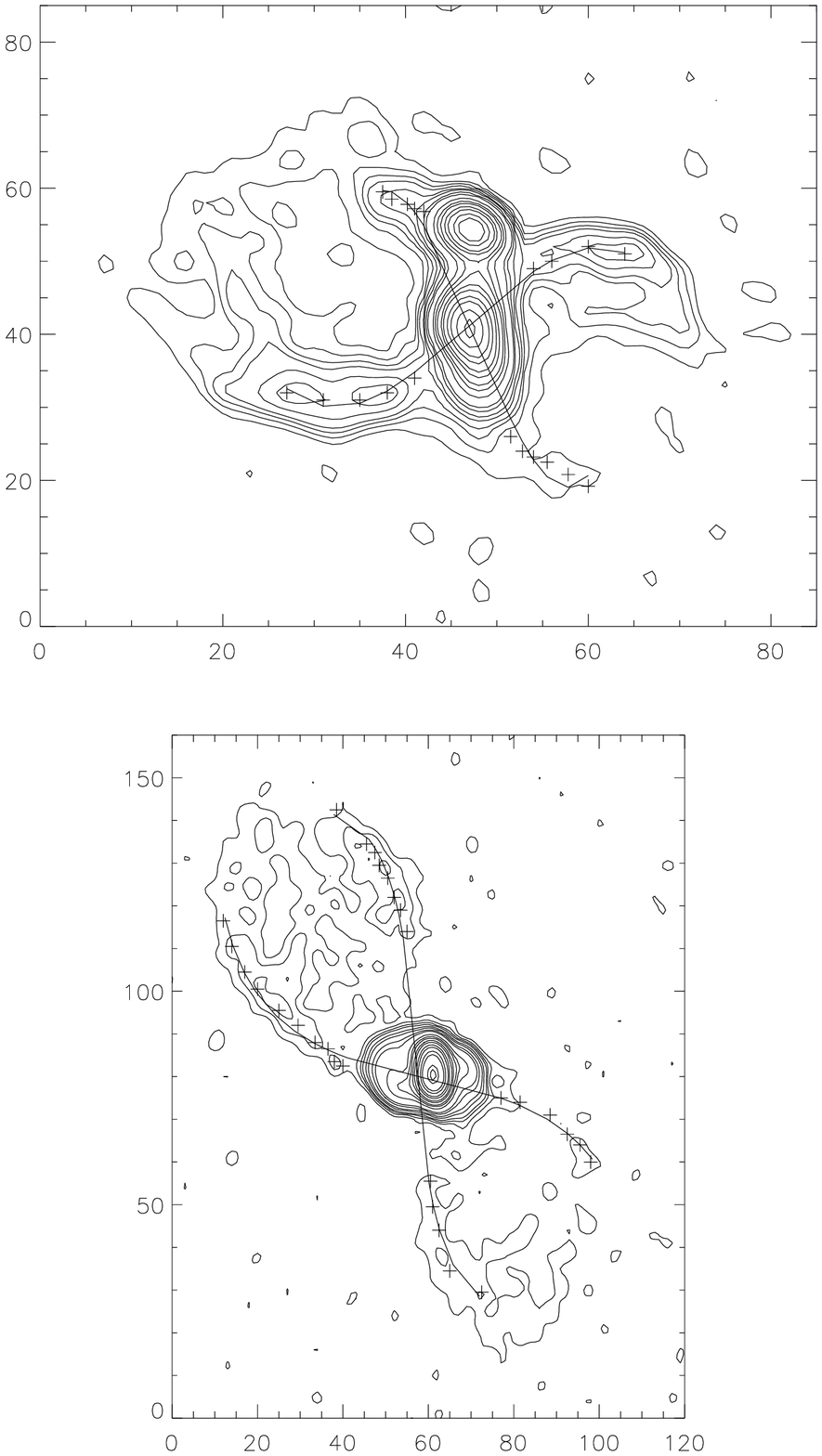}}
\caption{Extrapolations of bubbles down to their origins (see text).
The axes coordinates are in pixels, the pixel size being $0.1\arcsec$ in the
top and $0.25\arcsec$ in the bottom panel, respectively.
This figure indicates that both pairs of bubbles originate from
a region within 170 pc.}
\label{figrelic}
\end{figure}

In order to estimate the originating point of the bubbles (where the two edges 
meet, see Fig.~\ref{figrelic}), under the simplifying assumption that the two 
edges of a bubble on either side lie in the same plane, we used a simple 
IDL program which, 
given the pixel positions of certain bright spots on the edges of the bubbles, 
fits a third degree polynomial to them, and derives the point where 
the difference in the position of the fitted functions is a minimum. Thus, it 
turns out that the outer pair of bubbles originate near $\alpha$=06:52:12.359, 
$\delta$=+74:25:37.26 (J2000) while the inner pair of bubbles originate 
near $\alpha$=06:52:12.340, $\delta$=+74:25:37.081 (J2000). Both originate 
from near the same place (offset by $0\arcsec.3$ or $\sim$110 pc), which
interestingly, lies around the hot spots 3 and 4 in \citet{Kukula96} 
(to the south of the (0,0) position in the MERLIN image of 
Fig.~\ref{figmorph}). It should be pointed out that the bright spots we pick 
on the edges of the bubbles are not without ambiguities, which in turn give 
uncertain positions for the expected origins. The offset in the originating 
point of the bubbles may therefore not be of much significance.
As \citet{Kukula96} have noted, no radio `core' has 
been found coincident with the centre of the optical host galaxy, which
lies close to the (0,0) position (with a $3\sigma$ uncertainity in position of 
$0.27\arcsec$, \citet{Clements81}) in the MERLIN image (Fig.~\ref{figmorph}), 
and it is not clear if one of the knots $\sim0.5\arcsec$ to the south, 
is the radio core. 
\clearpage
\clearpage
\begin{table}[h]
\begin{center}
\caption{\label{tabenergetics} Physical Properties and Energetics}
\begin{tabular}{lcccccccccccccc} \hline \hline
Structure, Filling & Length& Width& L$_{rad}$  & E$_{min}$ & B$_{min}$  & P$_{min}$ \\
Factor ($\phi$)&($\arcsec$)&($\arcsec$)&(erg~s$^{-1}$) & (erg) & ($\mu$G) & (dynes~cm$^2$)\\
\hline
Outer Bubble, $\phi$=1&&&& \\
North       &20.4&12.1& 7.4E38 & 9.7E54 & 3.6 & 1.2E-12\\
South       &17.8&12.4& 3.8E38 & 6.4E54 & 3.0 & 8.7E-13\\
Inner Bubble, $\phi$=1&&&& \\
East        &3.8&3.7& 9.4e38 & 2.0E54 & 12.3 & 1.4E-11\\
West        &3.0&3.2& 4.2e38 & 9.9E53 & 11.3 & 1.2E-11\\
Outer Bubble, $\phi$=10$^{-3}$&&&& \\
North       &...&...& 7.4E38 & 5.1E53 & 25.9 & 6.2E-11\\
South       &...&...& 3.8E38 & 3.3E53 & 22.0 & 4.5E-11\\
Inner Bubble, $\phi$=10$^{-3}$&&&& \\
East        &...&...& 9.4E38 & 1.0E53 & 88.4 & 7.2E-10\\
West        &...&...& 4.2E38 & 5.1E52 & 81.4 & 6.1E-10\\\hline
North-South Jet   &3.1&1.0& 9.0E39 & 2.1E54 & 52.3 & 2.5E-10\\
Outer Jet (North) &6.8&3.0& 1.3E38 & 6.9E53 & 6.7  & 4.1E-12\\
                  &7.5&3.1& 1.5E38 & 8.1E53 & 6.7  & 4.1E-12\\
Outer Jet (South) &7.2&3.1& 8.7E37 & 5.7E53 & 5.7  & 3.0E-12\\
                  &5.4&2.0& 3.6E37 & 2.1E53 & 6.2  & 3.6E-12\\
Inner Jet (East)  &1.8&0.8& 3.5E38 & 2.2E53 & 27.4 & 6.9E-11\\
                  &1.5&0.7& 7.3E37 & 7.4E52 & 19.9 & 3.7E-11\\
Inner Jet (West)  &1.4&0.6& 3.2E37 & 3.9E52 & 17.6 & 2.9E-11\\
                  &1.5&0.7& 1.8E38 & 1.2E53 & 25.8 & 6.2E-11\\
\hline 
\end{tabular}
\end{center}
\tablecomments{Col.2 and Col.3: The length and width in arcsecs of the volume element 
considered for the calculations, using the 6~cm image.
$L_{rad}$ = averaged total radio luminosity (see Sect.~\ref{energetics}),
E$_{min}$ and B$_{min}$ correspond to the particle energy
and magnetic field strength at the minimum pressure P$_{min}$, obtained
assuming the `equipartition' condition. The upper part of the table has the
parameters listed for the bubbles which are either completely filled 
($\phi$=1) or mostly empty ($\phi$=10$^{-3}$), while the lower part of 
the table lists the parameters for the radio jet and volume elements
taken along the edges of the bubbles, assuming $\phi$=1. In the 
precessing jet scenario, these volume elements correspond to portions
of the precessing radio synchrotron jet. Outer Jet (North) corresponds to a
volume element taken along the edge of the northern outer bubble on 
either side, and so on.}
\end{table}
\clearpage
\subsection{Mechanisms for Producing Bubble-like Structures}
\label{secmech}

Extended radio structures have been observed in both Seyfert and starburst 
galaxies, and some of the structures are bubble-like \citep{Colbert96,
Heckman93}. 
In starburst galaxies, the bubble-like radio structures have been suggested to 
form through the interaction between the starburst superwind and the ISM
\citep{Duric88,Heckman93}. In the case of Seyfert galaxies, \citet{Colbert96} 
found that the kpc-scale radio outflows are more consistent with diverted 
AGN-driven jets. But they didn't rule out the possibility that the outflows 
are starburst superwinds. In fact, indications of both AGN and starburst 
activity have been found in Seyfert galaxies like NGC~7469 
\citep{Genzel95}, NGC~3079 \citep{Veilleux94} and the Circinus galaxy 
\citep{Maiolino98}. 

Other scenarios have been proposed to explain bubble formation in 
galaxies. In order to explain the spherical shell-like structures in the 
radio galaxies Her~A and 3C~310, \citet{Morrison96} introduced a model wherein 
acoustic waves or weak shocks are excited along the jet axis of the AGN, in a 
pre-existing thermal galactic wind. These acoustic shells expand uniformly 
at the speed of sound and drift along with the wind. At a much later second 
stage, a new and much faster flow of relativistic plasma is energized by the 
AGN, which fills these shells with radio-emitting electrons, thus creating 
radio-bright bubbles. Alternately, \citet{Pedlar85} invoked a model of 
expanding radio-emitting plasmons which shock, compress and accelerate
the forbidden-line gas in Seyfert galaxies. Both these models require the 
bubbles to be continuously injected with relativitic particles from the AGN. 

Bubble-like structures observed in some planetary nebulae 
like Hb~5, \citep{Terzian00}\footnote{http://ad.usno.navy.mil/pne/gallery.html}
or $\eta$ Carina \citep{Nota95}, have been suggested to arise due to the
presence of orbiting binary companions \citep{Soker04}.
A binary black hole system could similarly be invoked to explain the bubble-like
structures in Mrk~6. In this picture, the primary black hole is spinning
and is responsible for the formation of the radio jet. The secondary black 
hole on the other hand, is not spinning but orbits the primary black hole in 
a plane which is perpendicular to the inner bubbles.
The quasi-sherical disk wind from the primary black hole forms an 
accretion disk around the secondary black hole. This in turn launches 
fast winds into the pre-existing ``density contrast'' 
between the equatorial and polar directions \citep[see][]{Balick87} 
created by the orbiting secondary black hole.

The complexity involved in this binary blackhole model with the two backholes
having perpendicular orbital planes, however, makes it unattractive. For 
example, the orbital plane of the secondary black hole will be aligned with 
the inner jet axis and must exert a disturbing influence on its propagation 
\citep{Merritt05}, an effect which is not observed in the high resolution 
MERLIN image. Further, the existence of a secondary black hole orbiting in a 
plane perpendicular to the accretion disk of the primary, must have 
discernable effects \citep{Sillanpaa88,Lehto96} on the dynamics of 
the 100 pc-scale disk \citep{Gallimore98} which we assume to be relatively 
stable according to the ENLR \citep[][see Fig.~\ref{figwfpc}]{Meaburn89},
and the optical polarization position angle derived through spectropolarimetric
observations \citep{Smith04}. It is also uncertain whether the planetary 
nebula model will scale to an AGN.
In the following sections, we discuss in some detail the pros and cons of
three potentially viable
scenarios that could give rise to the bubble-like structures in Mrk~6,
$viz.,$ (i) a starburst-driven bubble scenario, (ii) a jet-driven bubble
scenario, and (iii) a precessing jet scenario.

\subsubsection{Starburst-driven Bubble Scenario}
The morphology of the pairs of the bubbles, 
especially their edge-brightening suggests that the radio emission could 
come from the shells of superbubbles. The sharp and well-defined edges of the 
bubbles and the existence of filaments in the bubbles would be consistent with 
strong interaction between the bubbles and the interstellar medium (ISM) 
where the radio-emitting particles (relativistic electrons) are accelerated 
and/or the magnetic field gets amplified \citep{Pacholczyk76}. 
For example, if the expansion of the bubbles are caused by diffusion
with little interaction (shocks), we would expect the bubbles
to be more irregular and diffusive, as seen in most radio lobes.
The spectral index of the bubbles seems rather flat, also
consistent with interaction and particle re-acceleration.

Assuming that the outer bubbles are formed through starburst winds energized 
by supernovae explosions, we estimate the energy budget. If we assume 
the bubbles are filled with the ejecta from the supernovae, the 
energy required for the bubbles to expand to the present sizes 
(V$\sim7.5\times4.5\times4.5$~kpc$^3$) adiabatically within the ISM of 
pressure P$\sim10^{-10}$ dynes~cm$^{-2}$ is 4PV=1.7$\times 10^{57}$ ergs. 
This corresponds to $1.7\times10^6$ supernova explosions with individual 
explosive energy of $10^{51}$ ergs. For wind speeds of 1000 km~s$^{-1}$, the 
time required for the bubbles to expand to the present size is 
$\sim7\times10^6$ yrs. Thus the required supernova explosion rate is 
$\sim$ 0.24~yr$^{-1}$, which is comparable to the predicted rate for the 
prototypical starburst galaxy M~82 \citep{Seaquist91,Rieke93}. The 
required supernova rate would be reduced if the bubbles are not fully 
filled with the superwinds. We must mention that we have not included the 
loss through radio emission since it is negligible compared with the 
dynamical energy we have estimated above. 

The derived supernova rate can yield an estimate of the required 
star-formation rate (SFR). By assuming a Salpeter initial mass function 
over the mass range 
$0.1\leqslant M/M_\sun\leqslant 100$, 
using the above derived supernova rate, we obtain
(by using Eqns. 28, \citet{Condon02} and 20,~\citet{Condon92}), a large 
star-formation rate of $\sim33~M_\sun$~yr$^{-1}$ for Mrk~6.
The SFR estimated from the 1.4 GHz luminosity of
the bubbles ($\sim6\times10^{22}$ W Hz$^{-1}$) for $\alpha=-0.6$ and 
Eqns. 21 of \citet{Condon92} and 28 of
\citet{Condon02}, turns out to be much larger 
(SFR $\sim$76 $M_\sun$~yr$^{-1}$). 

An upper limit to the SFR can be obtained from recent Spitzer
IRAC observations of Mrk~6 \citep{Buchanan06}. Using the
8$\micron$ flux density of 0.1 Jy resulting from the central $\sim7\arcsec$
region (corresponding to a spatial scale of 2.7 kpc), and the relation, 
SFR~=~2.0$\times10^{-43}~\nu~L_{\nu}(8~\mu m)$ erg~sec$^{-1}$,
(Calzetti et al. 2006, in prep.), which makes use of the 
nebular lines calibration of \citet{Kennicutt98},
we obtain a SFR ($M\geqslant 0.1 M_\sun$) $\sim$ 5.5 $M_\sun$~yr$^{-1}$. 
This value is much lower than the SFR derived assuming that the bubbles are 
a result of starburst winds energized by supernovae explosions.

The derived minimum pressure estimate for the bubbles 
($\sim10^{-11}-10^{-12}$~dynes~cm$^{-2}$) is one to two orders of 
magnitude lower than the typical ISM pressure of $\sim10^{-10}$~dynes~cm$^{-2}$ 
found in an elliptical or an S0 galaxy \citep{Mathews03}.
This could imply that either (i) the pressure has a large
contribution from thermal particles, (ii) the equipartition 
condition does not exist in the bubbles, or (iii) 
%the bubble-like structures are not actual bubbles but just the projected jet, 
the equipartition condition exists
%, the radio structures are in pressure equilibrium, 
and the derived pressure estimates are the actual ISM pressures in Mrk~6.
Further, through optical spectroscopic observations, \citet{HeckmanArmus90} 
have
found typical kpc-scale pressures of $2-4\times10^{-9}$~dynes~cm$^{-2}$, in 
a sample of galaxies with starburst-driven galactic superwinds.
The derived P$_{min}$ values for Mrk~6 are two to three orders of magnitude 
lower (see Table~\ref{tabenergetics}). 
If we assumed that the bubbles were powered by superwinds, then this
low P$_{min}$ could imply that there is a large thermal contribution
to the pressure, or the equipartition condition does not hold, or the bubbles 
are not powered by typical superwinds.
In their work on the X-ray binary SS433, \citet{Blundell01} have 
suggested that powerful accretion disk winds, which arise perpendicular to
the radio jet, can also contribute to the
radio luminosity. An accretion disk outflow cannot be ruled out in Mrk~6.

In summary, the superwind model has the advantage that the complex geometry 
of the radio emission and the different features can be understood as being 
determined by their own local environments. The disadvantage of the model is 
that it cannot explain the similarities between the two pairs of the bubbles, 
if the similarities are not simple coincidences. However, the most
important concern 
for this model stems from the derived energetics -- 
the upper limit to the SFR provided by Spitzer observations indicates
that a starburst alone cannot meet the energy requirements for the
creation of the bubbles in Mrk~6.

\subsubsection{AGN-driven Bubble Scenario}
The distorted appearance of many Seyfert jets has lead to the suggestion
that the jet diverts due to interaction with molecular gas, within the central
hundred parsecs \citep[e.g.,][]{Gallimore96,Bicknell98,Gallimore06}. 
The jet eventually terminates 
in a bow shock which plows into the surrounding ISM, resulting in 
bow-shock-like structures on kiloparsec scales, similar to the structures 
observed in Mrk~6. Assuming that the efficiency ($\epsilon$) with which the 
total jet energy is tapped to produce radio luminosity is 1\% 
\citep[see for e.g.][]{Odea85}, the jet with a radio luminosity of 
$\sim10^{40}$~ergs~sec$^{-1}$ (see Table~\ref{tabenergetics}), and a 
kinetic luminosity ($\frac{L_{rad}}{\epsilon}$) 
of $\sim10^{42}$~ergs~sec$^{-1}$,
would require $\sim3\times10^{5}$~years to deposit all its energy into the 
nuclear ISM and produce the derived minimum energy of $\sim10^{55}$ ergs 
%(Table~\ref{tabenergetics}) 
for the outer bubble. The jet would require only $\sim7\times10^{4}$~years 
to produce the derived minimum energy for the inner bubble.
Although this model seems energetically feasible it does not naturally
explain the two edge-brightened nested radio structures or the fact that
the inner jet is perpendicular to the inner set of bubbles.

We note that the pressure-volume energy of the outer bubble is around a 
hundred times larger than the derived minimum energy. Therefore, if the
equipartition condition holds, the energy contribution from the thermal 
particles to the bubbles, must be a hundred times greater than from the 
relativistic electrons.

AGN-driven plasmons have also been proposed to explain bubble-like structures
in Seyfert galaxies \citep{Pedlar85}. Models of radio emission from 
idealized plasmons \citep[e.g.,][]{Shklovskii60} suffer severe adiabatic losses 
resulting in their radio luminosity decreasing as $r^{-5}$, where $r$ is the 
radius of the plasmon. Assuming adiabatic losses in a spherically expanding 
radio plasmon of relativistic gas, the final luminosity and internal energy 
of the plasmon is given by $L=L_0(r_0/r)^5$ and $E=E_0(r_0/r)$, respectively,
where $L_0$ and $E_0$ are the initial luminosity and internal energy of 
the plasmon with initial radius $r_0$. Using 
$L=7.4\times10^{38}$~ergs~sec$^{-1}$ and $E=5\times10^{53}$~ergs at a 
radius $r$=7.5/2=3.75~kpc for the outer bubble (see Table~\ref{tabenergetics} 
for $\phi=10^{-3}$), we derive a large initial luminosity and internal energy 
of $L_0\sim6\times10^{56}$~ergs~sec$^{-1}$ and
$E_0\sim2\times10^{57}$~ergs, at the radius $r_0$=1~pc. Similarly for the inner 
bubble with $r$=3.8/2=1.9~kpc, we derive 
$L_0\sim2\times10^{55}$~ergs~sec$^{-1}$ and $E_0\sim2\times10^{56}$~ergs, at 
$r_0$=1~pc. 
However, typical bolometric luminosities for Seyferts galaxies 
fall in the range of $10^{45}-10^{46}$~ergs~sec$^{-1}$ \citep{Padovani88},
making the $L_0$ estimate unreasonable. Alternatively,
an energy equivalent of $10^5$ years of bolometric luminosity must 
be released at the radio frequency alone, at one instant. 
A single plasmon model for the bubbles of Mrk~6 therefore seems to be
energetically implausible.

\subsubsection{Precessing Jet Scenario}
In this model, an episodic precessing jet is the prime mover that has 
produced all of the radio structure visible in Mrk~6. Assuming a jet velocity 
of $\sim10^4$~km~s$^{-1}$ \citep{Bicknell98,Ulvestad99}, and the bubbles-edges 
to be a precessing jet, the outer bubbles would have resulted in
$\geqslant 7\times10^5$ years and the inner bubbles would need 
$\geqslant 1\times10^5$ years to form. The gap between the size-scales of the two 
pair of bubbles would imply different episodes of ejection separated by
$\geqslant 6\times10^5$ years. Therefore, the outer bubbles would be the 
relic of the radio jet emission from about 10$^6$ years ago as it precessed 
in the northeast-southwest direction, probably following an accretion event. 
The precessing jet subsequently changed direction and pointed in the east-west 
direction about 10$^5$ years ago, possibly following another accretion/ejection 
episode. 
If the jet velocity is closer to $10^3$~km~s$^{-1}$ \citep[e.g.,][]{Whittle04},
the time-scales for the formation of the outer and inner ``bubbles'' 
would be $\geqslant 7\times10^6$ years and $\geqslant 1\times10^6$ years, respectively.

\begin{figure}
\centerline{
\includegraphics[width=8cm]{f6a.ps}
\includegraphics[width=9cm]{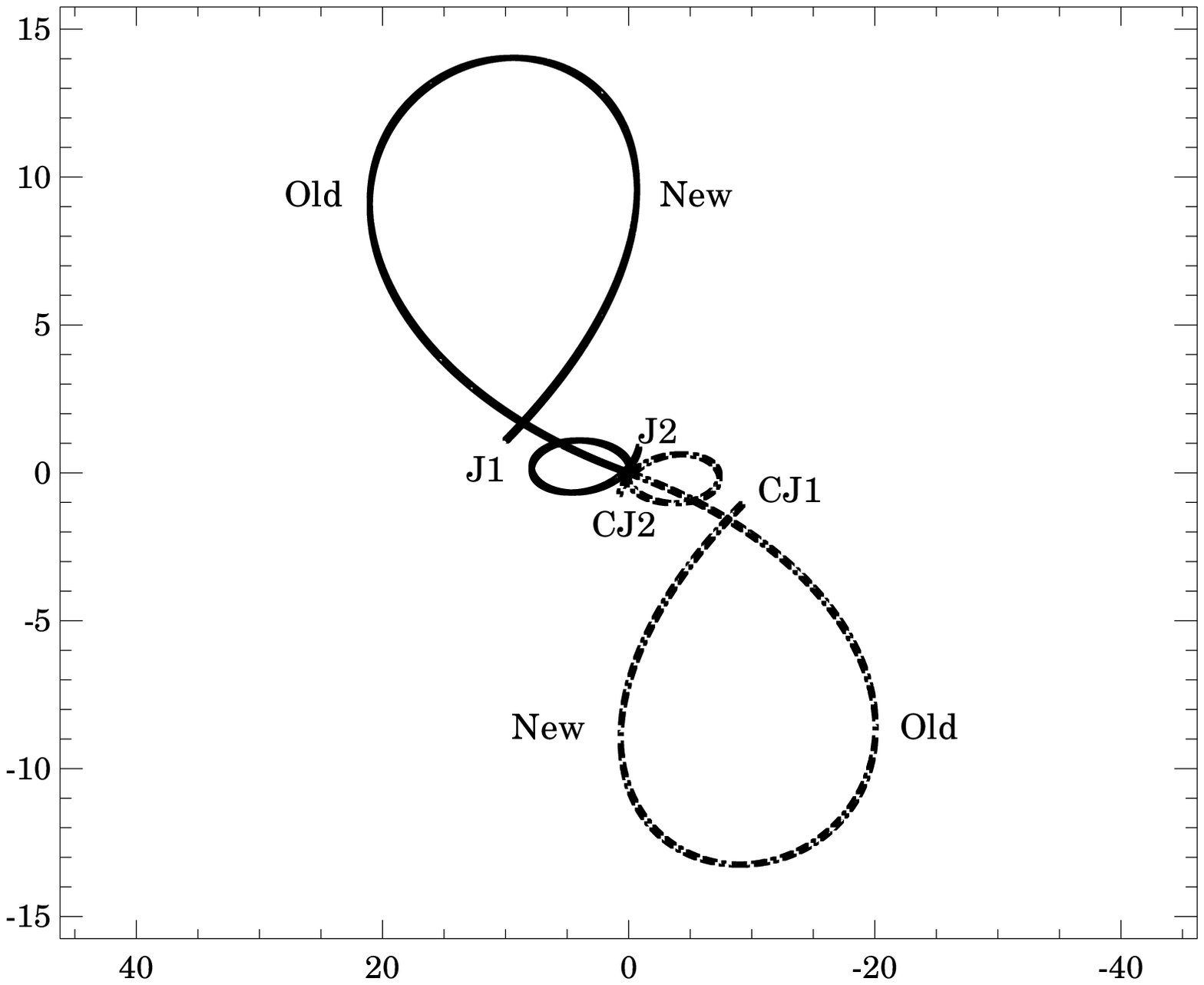}}
\centerline{
\includegraphics[width=8cm]{f6c.ps}
\includegraphics[width=9cm]{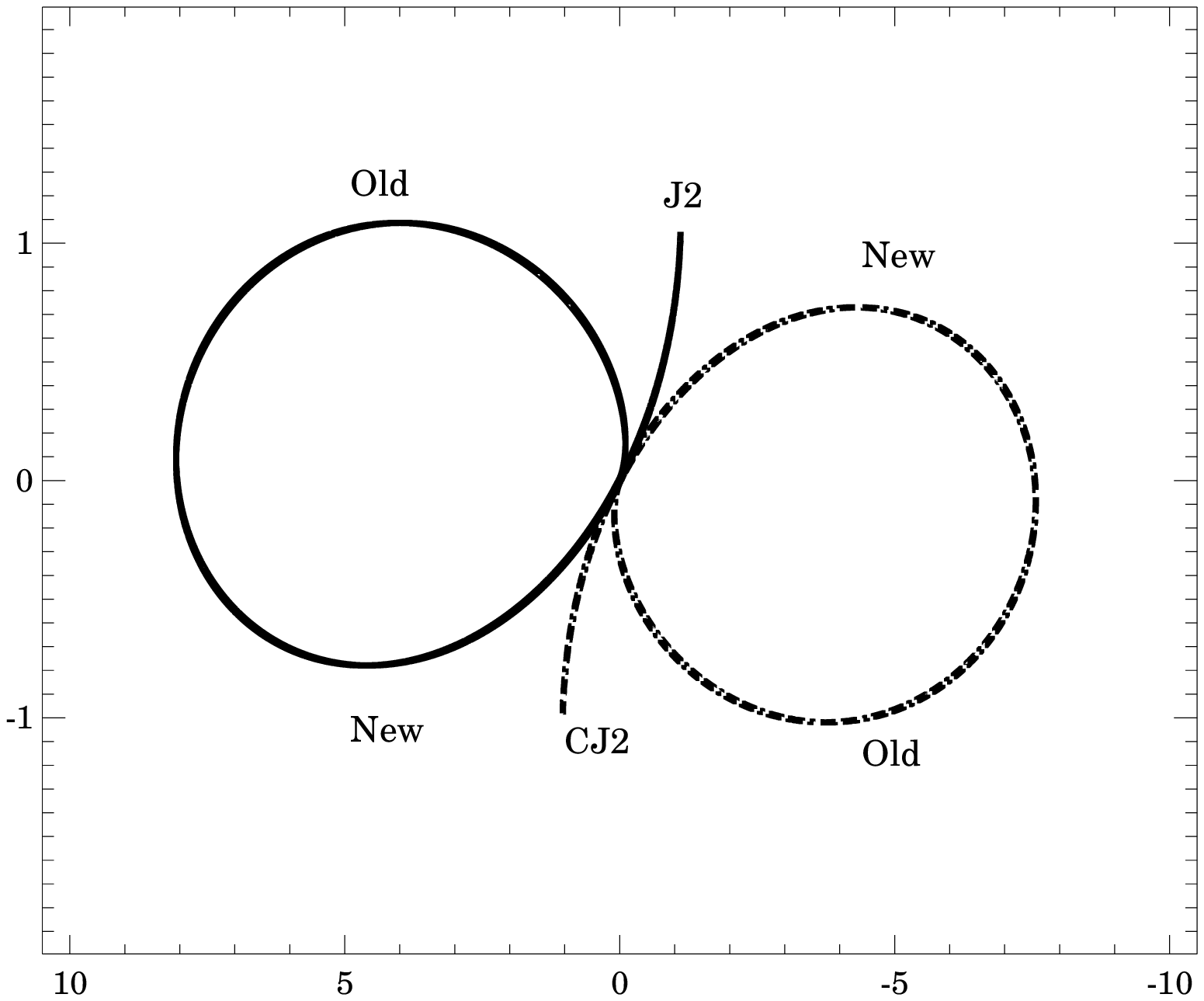}}
\caption{\small A comparison of the morphology of an episodically-powered 
precessing jet (top right) and the 20~cm radio image (top left). (Right)
Plots of the proper motions of the precessing jet projected on
the sky (\ie, the $\mu_\alpha$ and $\mu_\delta$ parameters as described in 
Hjellming \& Johnston, 1981). The axes are in arcseconds, 
the solid line represents the approaching jet, while the dashed line represents
the receding jet. The outer and inner ``bubbles'' are jets from two different
phases of activity of the AGN. J1, CJ1, J2 and CJ2 indicate the jet and 
counterjet from the two phases and their approximate end-positions.
In the precession model, the right edge of the northern outer 
bubble and the left edge of the southern outer bubble (both marked ``New'') 
would be brighter and have a flatter spectral index, as it 
is corresponds to the most recent position of the jet. The same applies to 
the bottom edge of the eastern inner bubble and the top edge of the western 
outer bubble (bottom panel). This agrees with the image on the left.
Finally, the S-shaped MERLIN jet is easily reproduced in this model.}
\label{figprec}
\end{figure}

With this picture in mind, we derived the proper motions of
two precessing jets with $\beta=0.03$, corresponding to the two
phases of activity of the AGN, using Eqns. 1--4 of \citet{Hjellming81}.
In Figure~\ref{figprec}, we have plotted the proper motions of the precessing 
jet projected on the sky (\ie, the $\mu_\alpha$ and $\mu_\delta$ parameters as
described in Hjellming \& Johnston, 1981), 
for two different jet inclinations and precession cone opening angles.
The parameters pertaining to this model are listed in 
Table~\ref{tabparams}. The model suggests that the northern outer 
and the eastern inner jets are approaching jets (denoted by solid lines
in Fig.~\ref{figprec}), while the southern outer
and the western inner jets are receding jets (denoted by dashed lines
in Fig.~\ref{figprec}).
Further, the outer jets precess in the clockwise sense, while the inner jets
precess in the counterclockwise sense. The jet flips its axis by 
$\sim140$ degrees between the two epochs. 

The similarity between the precessing jet model and the radio images is 
striking. In the model, assuming only adiabatic losses and minimal radiative 
losses in the radio emitting plasma, the right edge of the northern outer 
``bubble'' and the left edge of the southern outer ``bubble'' would both be 
brighter and have a flatter spectral index, as it would correspond to the
most recent position of the jet. Elsewhere the spectra would have steepened
due to the loss of energy in the synchrotron-emitting particles.
The same would apply to the bottom edge of 
the eastern inner ``bubble'' and the top edge of the western outer ``bubble''. 
As we see in Fig.~\ref{figprec} and Fig.~\ref{figspectr}, this is in good 
agreement with the observations. Further, the precession model easily 
reproduces the elongated morphology of the outer bubbles and the north-south 
S-shaped MERLIN jet -- it is the projected terminal 
position of the jet that formed the inner ``bubble'' structure.

The precessing jet model yields precession periods of $3\times10^6$ and
$4\times10^5$ years for the outer and inner jets, respectively
(Table~\ref{tabparams}). Further, the jets were ``on'' for approximately one 
precession period in each episode of activity. The similarity in the 
precession and the ``activity'' period is perhaps suggestive of a link 
between the two phenomena -- the accretion event itself could be the cause of 
the precession. It is very interesting that \citet{NatarajanPringle98} find the
timescale for the accretion disk/blackhole alignment 
for a $10^6-10^7 M_\sun$ blackhole 
\citep[see for example][for Seyfert blackhole estimates]{Peterson04} accreting 
at one-tenth of the Eddington limit, 
to be of the order of $6-7\times10^5$ years.
\clearpage
\clearpage
\begin{table}
\begin{center}
\caption{\label{tabparams} Precessing jet parameters}
\begin{tabular}{cccccccc} \hline \hline
Structure&$i$&$\psi$&$\chi$&$\beta$&$P$&$t$\\
&(deg)&(deg)&(deg)&&(yr)&(yr)\\ 
(1)&(2)&(3)&(4)&(5)&(6)&(7)\\\hline
Outer jet&153&24&--30&0.03&3$\times10^6$&1.7$\times10^6$\\
Inner jet&9&9&--30&0.03&4$\times10^5$&4$\times10^5$ \\ \hline 
\end{tabular}
\end{center}
\tablecomments{Col.1: In the precessing jet model, the outer and inner jets correspond to the
the outer and inner ``bubbles'' in Mrk~6,
Col.2: Inclination angle of the jet axis, Col.3: Half-opening angle of the
precession cone, Col.4: Angle needed to rotate the geometrical model so 
that the axes coincide with the true north and east \citep[see][]{Hjellming81},
Col.5: Jet speed with respect to speed of light, Col.6: 
Precession period in years, Col.7: Total time the jet was ``on'' in years,
assuming the jet speed given in Col.5.
The jet sign parameter, $s_{jet}=+1$ was used for the approaching northern jet
and $s_{jet}=-1$ for the receding southern jet. The rotation sign parameter, 
$s_{rot}=-1$ was used to imply clockwise motion of the precessing outer jet
while $s_{rot}=+1$ was used for the counterclockwise precessing inner jet.}
\end{table}
\clearpage
In order to test the precession model, we derived a spectral age using the
difference in the spectral index values taken from the two opposite edges 
of the eastern 
inner ``bubble'' (marked `New' and `Old' in Fig.~\ref{figprec}, bottom right),
under the assumption that the spectral steepening is a result of ageing.
Using the \citet{JaffePerola73} model for spectral ageing and following
the analysis described in \citet{MyersSpangler85}, we derived a radiative 
age of $4\times10^5$ years, assuming an equipartition magnetic field of
$\sim90\mu$G (Table~\ref{tabenergetics}). 
This estimate is a good match to the timescale the inner jet was 
``on'' in the precessing jet model, providing further support to the 
model.

The fractional polarization values along the edges of the bubbles range from
20\% to 40\%, values that are typically observed in AGN jets.
Typical values for the measured polarization of supernova remnants, on the
other hand, are less than 10\% on all scales
\citep[eg.,][]{Gaensler03,Lazendic04,McConnell06}. 
The polarized emission indicates that the southern side of the outer and the
western side of the inner bubbles is more polarized than the opposite sides. 
%which could potentially clash with the orientation picture outlined above. 
However, the radio polarization seems to be anticorrelated with the presence 
of the emission-line gas \citep[see][]{Kukula96}, suggestive of depolarization 
of the northern and eastern bubble by ionized gas. Such an 
anticorrelation has also been observed in some radio galaxies 
\citep{Heckman82,LiuPooleyA91}. 
We have estimated P$_{min}$ for volume elements along the edges of the bubbles,
assuming a volume filling factor of unity (see Table~\ref{tabenergetics}).
In the precessing jet scenario, these correspond to sections of the 
radio synchrotron jet. We find the minimum pressure to be 
$\sim3-6\times10^{-11}$~dynes~cm$^{-2}$ and the minimum pressure magnetic field
to be $\sim16-26~\mu$G for sections of the inner bubble edge. 
The minimum pressure and magnetic field strengths 
are lower by an order of magnitude in portions of the outer bubble edge.
We find that these values are similar to those obtained for radio jets in 
radio galaxies \citep{BridleChan81,Bridle84} and Seyferts 
\citep[\eg,][]{Pedlar85}. 

The 6~cm MERLIN image of Mrk~6 reveals a gentle S-shaped knotty radio source 
(easily reproduced in our precession model, see Fig.~\ref{figprec}) which 
has often been suggested to be the signature of a precessing jet. 
%The steeper spectral index of this jet 
%compared to the rest of the radio structure could then suggest that the 
%precessing radio jet terminates and fades away at this position.
Episodic ejection of material into a two-sided, precessing (or wobbling), 
collimated outflow or jet has been proposed to explain the point-symmetric 
morphology of planetary nebulae \citep{LivioPringle97,Raga93,Lopez93,Soker94,
LivioPringle96}. Specific examples include the pre-planetary nebulae IRAS 
16342-3814 \citep{Sahai05} and Hen 3-1475 \citep{Velazquez04}, the planetary 
nebulae K3-35 \citep{Miranda01} and M2-9 \citep{Schwarz04}, the AGB star 
W43A \citep{Imai02}, the symbiotic star CH Cygni \citep{Crocker02} and 
Herbig-Haro type jets \citep{Lim01}.

The S-shaped 
symmetry in jets has been observed in Seyfert galaxies like Mrk~3, 
NGC~4151, NGC~5256, NGC~2685 and radio galaxies like 3C~305, 3C~345, 3C~294, 
3C~334, and 3C~120 \citep{Kukula93,Pedlar93,Heckman82,CaproniAbraham04,
Caproni04}. Most Seyfert galaxies observed on arcsecond-scales, do not
show extended jets \citep[\eg][]{Nagar99}. However when jet-like features
are observed, they more often than not, exhibit a curved geometry 
\citep[\eg][]{UlvestadWilson84B,Kukula95,Nagar99,Thean00}.
Indeed, 8 or S-shaped radio structures, sometimes along with large 
misalignments between
radio structures on different spatial scales have been observed in the
Seyfert galaxies NGC~2992, NGC~3079, NGC~5548, NGC~4051, NGC~6764, NGC~2110,
NGC~4151, NGC~7469 \citep{WehrleMorris88,Duric88,Baum93,Nagar99,Thean01}.

Further support for the precessing jet model comes from the emission line 
images obtained with the Hubble Space Telescope \citep{Capetti95}, where 
the [OIII] and [OII] emission of Mrk~6 follows the curve of the S-shaped 
jet far into the southern edge of the inner bubble. 
It would be difficult to explain this assymmetry in emission-line morphology
with an expanding shell of starburst superwinds. 
The entire narrow-line region itself extends much
beyond the extent of the radio source. A similar curvature in the narrow-line 
region and an ENLR much larger than the radio
source have been observed in the Seyfert 2 galaxy, Mrk~3 \citep{Kukula93} and 
the radio galaxy 3C~305 \citep{Heckman82}. Similar to the suggestions made
for these galaxies \citep[see][]{Kukula93}, we interpret that the small-scale
emission-line region is powered by the radio jets while the ENLR
may be ionized by the radiation from the nucleus. 
However, in keeping with the picture of an episodic
precessing jet, it is possible that some of the emission in the ENLR
is a result of an earlier interaction with the radio jet.

Therefore, based on all the above findings, we believe that the complex 
radio morphology of Mrk~6 can be 
understood in terms of an episodically-powered precessing jet. 
Mrk~6 may therefore be a unique object only in the sense that we are able 
to observe the source in its short-lived phase of episodic activity and before
its radio emission from the previous epoch has faded away. 
Hence, similar radio structures may be 
found in other Seyfert galaxies going through such a phase,
and when they are observed with sufficient resolution and sensitivity
to image the relic emission from an earlier phase of activity.

\subsubsection{Implications of an Episodic Precessing Jet}

An episodically-powered precessing jet model in Seyfert galaxies
implies that radio activity is a short-lived phenomenon in the lifetime 
of a Seyfert galaxy. Further, multiple episodes of ejection are possible. 
These might be caused by independent accretion events. Based on the fraction 
of spiral galaxies which are Seyferts, \citet{Sanders84} estimated the 
statistical Seyfert lifetime to be of the order of $3-7\times10^8$ yrs. 
Further, based on the physical extents of the narrow emission
line regions and radio jets, Sanders concluded that Seyfert nuclear activity 
must be short-lived stochastic accretion events with no single episode lasting 
longer than $10^6$ yrs. Although the estimate of the fraction of Seyferts
in spirals has increased from a few percent to $\sim10\%$ following more 
recent surveys \citep{Ho97}, thereby increasing the statistical lifetime 
to $10^9$ yrs, the episode lifetime remains the same.
Random jet orientations in Seyferts occur, in his picture, 
because each accretion event is accompanied by its own independent accretion 
disk in no preferred plane. Based on the NICMOS imaging of a large sample of 
Seyfert galaxies, \citet{HuntMalkan04} also reached the conclusion that Seyfert 
activity may be prompted by a disturbance.

The precessing jet model further implies that the jet ejection axis is 
easily perturbed in Seyferts. An accretion disk which is irradiated by a 
central AGN can become unstable to becoming warped, resulting in the 
wobbling of the jet \citep{Pringle96,Pringle97,LivioPringle97} and severe 
misalignment of the jet axis and the axis of the outer accretion disk. 
Apart from radiative feedback that may warp a rotating accretion disk,
jet precession could arise due to near-Eddington accretion events that may 
alter the spin of the supermassive black hole, in turn affecting the 
orientation of the radio jet \citep{Rees78,ScheuerFeiler96,NatarajanPringle98}; 
binary blackholes \citep{CaproniAbrahamB04} and 
blackhole mergers, that might result in short timescale redirection of the jet 
axis \citep{Merritt02,Merritt05}. 
Possibly due to the smaller blackhole masses in Seyfert galaxies
compared to typical radio galaxies \citep[e.g.,][]{McLureDunlop01}, 
the spinning blackhole axis 
and consequently the radio jet axis is perturbed easily when accretion events 
take place, resulting in precessing jets and switches in jet-orientation. 

It must be noted that the host galaxy of Mrk~6 shows no conspicuous
signatures of galaxy-galaxy interaction or merger-activity, which could
lead to a binary black hole in the centre \citep[\eg,][]{Begelman80}. 
In fact, Mrk~6 seems to be located in a sparse environment 
\citep{Dahari85,Pfefferkorn01}, although the possibility of a major merger 
more than a dynamical timescale ($\sim10^9$ yrs) ago, cannot be ruled out.
Note that the emission-line region in Mrk~6 is extended along more than one
axis (Fig.~\ref{figwfpc}). 
However, as \citet{Dennett-Thorpe02} point out from their study on `winged'
radio galaxies, there remain two possibilities that could result in jet 
realignment without a second black hole. (1) An ingested dwarf galaxy may not 
leave observable traces that a larger merger would, and yet provide adequate 
mass and angular momentum to cause jet re-orientation. (2) The jet axis may 
not be determined
by the black hole alignment, but be strongly influenced by the accretion
disk. Instabilities in the inner disk could then cause switching of the jet
direction over short timescales.

A consequence of the changes in jet orientation is that the jet may 
interact with the molecular torus, stirring up the constituent matter. 
This would be consistent with the variable absorbing gas column density 
inferred through XMM-Newton X-ray spectroscopic observations of the central
regions of Mrk~6 by \citet{Schurch06}.   

\section{SUMMARY AND CONCLUSIONS}

We have carried out an extensive radio study with the VLA on the Seyfert 1.5
galaxy Mrk~6 and imaged a spectacular radio structure in the source.
The radio emission occurs on three different spatial scales,
from $\sim$7.5~kpc bubbles to $\sim$1.5~kpc bubbles and a $\sim$1~kpc radio jet,
all lying roughly orthogonal to each other. All the radio structures appear to 
originate from near the AGN core. 

We have considered three models to explain the complex radio morphology of 
Mrk~6. In the first model, we have suggested that the bubble-like radio
structures are the result of superwinds ejected by a nuclear starburst.
However, recent Spitzer observations of Mrk~6 provide 
an upper limit to the star formation rate of $\sim5.5~M_\sun$~yr$^{-1}$,
an estimate lower by an order of magnitude than the SFR of 
$\sim33~M_\sun$~yr$^{-1}$ 
derived assuming that the bubbles are a result of starburst winds energized 
by supernovae explosions. Thus, a starburst alone cannot meet the energy 
requirements for the creation of the bubbles in Mrk~6.

We have also considered a model wherein the bubbles are a result of diverted
AGN-driven jets. The small-scale radio jet gets diverted and decelerated 
through its interaction with the surrounding
interstellar medium, thereby depositing all its energy into the ISM in about
$10^5$ years. The energetics however, do not favour the AGN-driven single 
plasmon model proposed by \citet{Pedlar85}.

In the third model, we attribute the complex radio structure to be the 
result of an episodically-powered precessing jet that changes orientation. 
The attraction of this model is that it can naturally explain the complex radio 
morphology. It is also consistent with the energetics, the spectral 
index, and the polarization structure. Recent XMM-Newton X-ray observations
of a variable absorbing gas column density could be inferred as 
in-keeping-with the changing jet orientation in Mrk~6 which causes the jet
to interact with the torus, thereby stirring up the gas/dust.
Radio emission in this scenario is short-lived
phenomenon in the lifetime of a Seyfert galaxy which results due to an
accretion event. Further, possibly due to the smaller blackhole masses in 
Seyferts compared to typical radio galaxies, the spinning blackhole axis 
and consequently the radio jet axis is perturbed easily when accretion events 
take place, resulting in precessing jets and switches in orientation.
Multi-wavelength comprehensive radio studies of Seyfert galaxies would 
therefore reveal many more sources with similar complex radio structures 
as Mrk~6.

\section{ACKNOWLEDGEMENTS}
We thank the referee for insightful comments that improved the paper.
We thank Prof. Martin Elvis for a stimulating discussion on the paper.
The National Radio Astronomy Observatory is a facility of the National 
Science Foundation operated under cooperative agreement by Associated 
Universities, Inc.
This research has made use of the NASA/IPAC Extragalactic Database (NED) 
which is operated by the Jet Propulsion Laboratory, California Institute 
of Technology, under contract with the National Aeronautics and Space 
Administration.

\end{document}